\documentclass[namedreferences]{SolarPhysics}

\usepackage[optionalrh]{spr-sola-addons}
\usepackage{graphicx}
\usepackage{amssymb}       
\usepackage{color}         
\usepackage{url}           

\newcommand{\etal}{{\it et al.}}

\newcommand{\pder}[2]{ \frac{\partial #1}{\partial #2} }

\begin{document}

\begin{article}

\begin{opening}

\title{Calculation of Spectral Darkening and Visibility Functions for Solar Oscillations}

\author{C.~\surname{Nutto}$^{1}$\sep
        M.~\surname{Roth}$^{2}$\sep
        Y.~\surname{Zhugzhda}$^{1,3}$\sep
        J.~\surname{Bruls}$^{1}$\sep
        O.~\surname{von der L\"uhe}$^{1}$
       }
\runningauthor{C. Nutto \etal}
\runningtitle{Spectral Darkening and Visibility Functions for Solar Oscillations}

   \institute{$^{1}$ Kiepenheuer-Institut f\"ur Sonnenphysik, Sch\"oneckstra\ss e 6, 79104 Freiburg, Germany
                     email: \url{nutto@kis.uni-freiburg.de}\\ 
              $^{2}$ Max-Planck-Institut f\"ur Sonnensystemforschung, 37191 Katlenburg-Lindau, Germany \\
              $^{3}$ Institute of Terrestrial Magnetism, Ionosphere and Radio Wave Propagation of the Russian Academy of Sciences,              Troitsk City, Moscow Region, 142092, Russia
             }

\begin{abstract}

Calculations of spectral darkening  and visibility functions for
the brightness oscillations of the Sun due to global solar
oscillations are presented. This has been done for a broad range of the
visible and infrared continuum spectrum. The procedure for the calculations of these functions includes the numerical computation of depth-dependent derivatives of the opacity caused by {\it p} modes in the photosphere. A radiative transport code was used for this purpose in order to get the disturbances of the opacities due to temperature and density fluctuations. The visibility and
darkening functions are obtained for adiabatic oscillations
under the assumption that the temperature disturbances are
proportional to the undisturbed temperature of the photosphere. The
latter assumption is the only way to explore any opacity effects
since the eigenfunctions of {\it p} mode oscillations have not been obtained so far. 
This investigation reveals that opacity effects have to be taken into account since
they dominate the violet and infrared part of the spectrum. Due to this, the
visibility functions are negative for those parts of the spectrum. Furthermore, the
darkening functions show a wavelength dependent change of sign for
some wavelengths due to these opacity effects. However, the visibility and
darkening functions under the assumptions used are in contradiction
with observations of global {\it p} mode oscillations. But it is
beyond any doubt that the opacity effects influence the
brightness fluctuations of the Sun due to global oscillations.

\end{abstract}
\end{opening}

\section{Introduction}

Since the ACRIM experiment \cite{woodard84} it is well known that
solar oscillations lead to brightness fluctuations of the Sun as a
star. However, the spatial integration of the flux disturbances
over the whole solar disk suppresses high frequency modes and
photometric observations of the Sun as a star can only resolve
spatial oscillation modes with low harmonic degree ($\ell <3$). But if the
visibility of these low degree modes is suppressed at some
positions on the solar disk due to opacity effects, then it might
be possible to observe oscillation modes with harmonic degree
$\ell\geq 3$.

The space experiment IPHIR onboard the PHOBOS spacecraft
\cite{froehlich88} showed that the amplitudes of the brightness
fluctuations due to {\it p} modes are different for different optical
wavelengths. \inlinecite{zhugzhda93} showed with the
consideration of non-adiabatic waves in a uniform, non-grey
atmosphere that this problem cannot be solved with the
introduction of the Rosseland mean opacity. With the calculation
of visibility functions for low-degree non-radial oscillations and
a comparison with IPHIR data, \inlinecite{toutain93} suggested two
limiting cases for the explanation of the origin
of the brightness fluctuations:
\begin{itemize}
 \item intensity and flux perturbations, caused by adiabatic oscillations, are driven
 by opacity effects and thus have to be taken into account for calculations of visibility
 functions
\item the visibility functions depend on non-adiabatic effects and opacity disturbances
can be neglected (blackbody approximation).
\end{itemize}
The direct way to solve the problem is to solve the eigenproblem
for nonadiabatic oscillations for the solar model which includes
a standard model of the photosphere. To our knowledge nobody has done
this so far since the set of integro-differential equations of
radiative hydrodynamics has to be solved.
\inlinecite{staude94} and \inlinecite{zhugzhda96} proposed to use
another approach which assumes that the temperature and density
fluctuations due to {\it p} modes are given. Under some
special assumptions this makes it possible to obtain so-called darkening functions which
shows the dependence of brightness fluctuations with respect to
the position on the solar disk. The intention was to develop a
qualitative explanation of brightness oscillations in order to
improve observations of solar oscillations. The different
wavelengths under consideration indicated that the brightness
fluctuations of the most dominant oscillation mode ($\ell =0$) vanishes
for certain positions on the solar disk, {\it e.g.} the darkening
function changes its sign. With the assumption that these
calculations can be easily extended to modes with harmonic degree
$0<\ell<3$ and that those functions will also show a change of sign,
it is seen as an indication that modes with higher harmonic
degree, $\ell\geq 3$, might be observable.

\inlinecite{zhugzhda96} carried out calculations of darkening functions for a few wavelengths only. Besides, the
visibility functions were not calculated at all. In this paper the
calculations of darkening and visibility functions are
performed for a wide range of wavelengths, from the near UV to the 
infrared in order to see whether the opacity effects are essential 
for this range of the spectrum. This investigation will help to analyze the
observations from photometric instruments like DIFOS on board CORONAS-F \cite{lebedev04}.

\section{Intensity and Flux Fluctuations due to {\it P} Modes} \label{Sec:Theory}

To avoid the full solution of the equations of radiation hydrodynamics
\cite{zhugzhda96} we expand the general
expression for the intensity in a series around the equilibrium state up
to the first order of small disturbances of the opacity.
Assuming local thermodynamic equilibrium (LTE), the fluctuations
of the emergent intensity ($\delta I_\nu$) normalized to the
specific intensity ($I_{0\nu}$) at the center of the solar disk are
\cite{zhugzhda96}
\begin{eqnarray}
\frac{\delta I_\nu}{I_{0\nu}} &=& \int_0^\infty \mathrm e^{-\tau_\nu/\mu} \biggl[\frac{\mathrm d B_\nu(\tau_\nu)}{\mathrm d \ln T_0} \frac{\delta T}{T_0} + \frac{\delta\kappa_\nu}{\kappa_{0\nu}} (B_\nu(\tau_\nu)-I_\nu(\tau_\nu,\mu))\biggr] \frac{\mathrm d\tau_\nu}{\mu} \nonumber\\ && \times \biggl[\int_0^\infty \mathrm e^{-\tau_\nu} B_\nu(\tau_\nu) \mathrm d\tau_\nu \biggr]^{-1}\, , \label{Eq:relative_IntensFluct}
\end{eqnarray}  
where $\tau_\nu$ is the optical depth at the frequency $\nu$, $B_\nu$ the Planck function, $\mu= \cos\theta$, and $\theta$ the polar angle. The unperturbed, outgoing intensity, originating at optical depth $\tau_\nu$, is 
\begin{equation}
I_\nu(\tau_\nu, \mu) = \int_{\tau_\nu}^{\infty} \mathrm e^{-(\tau_\nu'-\tau_\nu)/\mu} B_\nu(\tau_\nu')\frac{\mbox{d}\tau_\nu'}{\mu}\, .\label{Eq:Int_0}
\end{equation} 
The fluctuation of the opacity ($\delta\kappa_\nu(\rho,T)$) is given by the fluctuation of the density ($\delta \rho$) and the temperature ($\delta T$),
\begin{equation}
\frac{\delta\kappa_\nu}{\kappa_{0\nu}} = \frac{\mbox{d}\ln \kappa_{0\nu}}{\mbox{d}\ln T} \frac{\delta T}{T_0} + \frac{\mbox{d}\ln \kappa_{0\nu}}{\mbox{d}\ln \rho} \frac{\delta\rho}{\rho_0}\ .\label{Eq:Fluctuation_Opacity}
\end{equation}  
Equation (\ref{Eq:relative_IntensFluct}) considers the various
contributions to the intensity fluctuations. The first term
appears because of the temperature fluctuation in the source
function, which can be described by the Planck function in the case of
LTE. The second term appears because of the perturbation of the
opacity due to disturbances in density and temperature. An approximation for the temperature and the density perturbations will be used.
The solution yields the intensity fluctuation of a non-grey atmosphere caused by small
disturbances. This can be used for the disturbances due to {\it p} modes
since they are small enough and the linear approximation works
perfectly.

To simplify Equation (\ref{Eq:relative_IntensFluct}) it is assumed that
disturbances of the equilibrium state are adiabatic. In this
special case the temperature and density disturbances are
connected by a simple relation:
\begin{equation}
\frac{\delta T}{T} = (\Gamma_3-1) \frac{\delta \rho}{\rho}\, ,  \label{Eq:adiabatic}
\end{equation}  
where $\Gamma_3$ is the third adiabatic exponent. For a neutral or
fully ionized gas it is constant and has the value $\Gamma_3 - 1 =
2/3$, whereas for a partially ionized gas it can reach values of
$\Gamma_3 - 1 \leq 0.1$. The first case is given for the
photosphere, which is completely dominated by neutral hydrogen. The
second case applies for the deeper photosphere and the
chromosphere where the gas is partially ionized.\newline 
In the case of adiabatic disturbances, the intensity fluctuations
depend only on temperature fluctuations and on parameters of the
quiet undisturbed photosphere. The dependence of the intensity
fluctuations on $\mu$ is a darkening function for the adiabatic
disturbances. To obtain the darkening function for {\it p} modes the
dependence of the temperature disturbances on depth ($\delta T(\tau_\nu)$)
has to be substituted into Equation (\ref{Eq:relative_IntensFluct}). But this function is not
known, even for adiabatic disturbances. 
In order to make a qualitative analysis of the opacity effects we simplify the
problem and it is assumed that the disturbances are proportional to
their undisturbed values \cite{zhugzhda96},
\begin{equation}
 \frac{\delta T}{T_0} = \mathrm{constant} \label{Eq:Temp_Perturbation}
\end{equation} 
which is a rather crude approximation. It holds only for continuum radiation which originates from a thin layer of the solar atmosphere. With this assumption the relative disturbances $\delta T/T_0$ and
$\delta \rho/\rho_0$ are independent of the optical depth
($\tau_\nu$) and can be factored out of the integral. Now, with
relationship (\ref{Eq:adiabatic}) the relative intensity variation
can be described in units of $\delta\rho/\rho$ or $\delta T/T$.
This leads to the darkening function
\begin{equation}
 \Delta_\nu=\frac{\delta I_\nu/I_{0\nu}}{\delta T/T_0}\, ,
\end{equation}  
which in our case is expressed in units of $\delta T/T_0$.

To obtain the visibility function, the darkening function has to be
integrated over the entire solar disk. This yields the following
expression for the flux perturbation for the most dominant mode
$\ell=0$:
\begin{eqnarray}
\frac{\delta F_\nu/F_{0\nu}}{\delta T/T_0} &=& \left\lbrace \int_0^{\infty}\biggl[\pder{B_\nu(\tau_\nu)}{\ln T} E_2(\tau_\nu)\right.\nonumber\\ 
& - & \biggl(\pder{\ln\kappa_\nu}{\ln T} \biggr)_\rho \int_{\tau_\nu}^{\infty} \frac{\mathrm{d} B_\nu(\tau_\nu')}{\mathrm{d}\tau_\nu'} E_2(\tau_\nu') \mathrm{d}\tau_\nu' \nonumber\\
& - & \left.\frac{1}{\Gamma_3(\tau_\nu)-1} \biggl(\pder{\ln\kappa_\nu}{\ln \rho} \biggr)_T \int_{\tau_\nu}^{\infty} \frac{\mathrm{d} B_\nu(\tau_\nu')}{\mathrm{d}\tau_\nu'} E_2(\tau_\nu') \mathrm{d}\tau_\nu' \biggr]\mathrm d\tau_\nu \right\rbrace \nonumber\\
&\bigg/& \left\lbrace \int_0^{\infty}B_\nu(\tau_\nu) E_2(\tau_\nu) \mathrm d \tau_\nu \right\rbrace \, . \label{Eq:relVisibil}
\end{eqnarray}

\section{Numerical Calculation of Darkening and Visibility Functions for Solar Oscillations with Degree $\ell=0$}

For the evaluation of the darkening functions one needs to know radiative transport quantities for the solar atmosphere as functions of the optical depth $\tau_\nu$. For this purpose the radiative transport code RH\footnote{\url{http://www.nso.edu/staff/uitenbr/rh.html}} by Han Uitenbroek was used \cite{uitenbroek01}. The code is based on the MALI (Multi-level Approximate Lambda Iteration) formalism by \inlinecite{rybicki91}. It solves the combined equations of statistical equilibrium and radiative transport under the general assumption of non-LTE conditions for a multilevel atom in a given plasma. In our case we choose hydrogen to be the considered atom. In addition to opacities and emissivities from the transitions in the hydrogen atom, the code accounts for background radiation sinks and sources due to other atoms, molecules, and all relevant continuum processes such as H$^-$ bound-bound and bound-free processes, Rayleigh scattering off neutral hydrogen and Thompson scattering off free electrons, helium, H$_2$ and hydrogen free-free processes. The considered atoms and molecules for the background radiation and opacities are listed in Table \ref{Tab:Backgroundelements}.
\begin{table}[h]
\begin{tabular}{ll}
\hline
Atoms & He, C, N, O, Na, Mg, Al, Si, S, Ca, Fe\\
\hline
Molecules & H$_2$, H$_2^+$, C$_2$, N$_2$, O$_2$, CH, CO, CN, NH, NO, OH, H$_2$O \\
\hline
\end{tabular}
\caption{Atoms and molecules that are considered as background elements by the radiative transport code.} \label{Tab:Backgroundelements}
\end{table}
In addition, it is necessary to provide the radiative transport code with a model of the solar atmosphere. We choose the semi-empirical one-dimensional quiet-Sun model FAL-C by \inlinecite{fontenla99}.

For the calculation of the darkening function we are not only interested in the radiative quantities alone, but also in their disturbances. Considering Equation (\ref{Eq:relative_IntensFluct}), we need the disturbances of the Planck function due to temperature variations and the disturbances of the opacity due to variation of temperature and density. \newline
In order to calculate the derivatives 
\begin{equation}
\biggl(\pder{B_\nu}{\ln T}\biggr)\,\mathrm{,} \quad \biggl(\pder{\ln \kappa_\nu}{\ln T}\biggr)_\rho\,\mathrm{,} \quad \mathrm{and} \quad \biggl(\pder{\ln \kappa_\nu}{\ln \rho}\biggr)_T \, , \nonumber
\end{equation}  
we disturb the quantities $T \mbox{ and } \rho$ of the atmospheric model which is then used as input for the radiative transport code. We use temperature disturbances of $-5\,$K and $+5\,$K, which is less then one percent of the undisturbed temperature of the atmosphere and the derivatives behave linearly under these small temperature disturbances. With Equation (\ref{Eq:adiabatic}) for the adiabatic relation, it can be assumed that the relative disturbances of the density are around 1\%. Thus, the hydrogen and electron populations are modified accordingly in the model of the atmosphere. While one modification is applied, all other values are kept constant. This procedure allows us to calculate the partial derivatives of the Planck function and the opacity using the centered-difference approximation \cite{numericalbook}.

Our interest is in the calculation of the darkening and visibility functions for a broad range of the visible and infrared spectrum. This has been done for 79 wavelength points between $250\,$nm and $2100\,$nm. We have to avoid any lines since we assume a temperature disturbance that is constant, see Equation (\ref{Eq:Temp_Perturbation}). This is only justified as long as the radiation  originates from a thin layer. This applies for the continuum radiation, but spectral lines form in higher layers and are not constrained to a small part of the solar atmosphere.

\section{Results}

\subsection{Derivatives}
First, the derivatives of the Planck function $(\partial B_\nu/\partial\ln T)$ were calculated.
The Planck function has the advantage that the derivative can be evaluated analytically and
the numerical values can be compared with the analytical function. A comparison showed that
the numerical calculation of the derivatives coincides with the analytical values within a
few tenths of a percent.

\begin{figure}
\centering
\includegraphics[width = 0.8\textwidth]{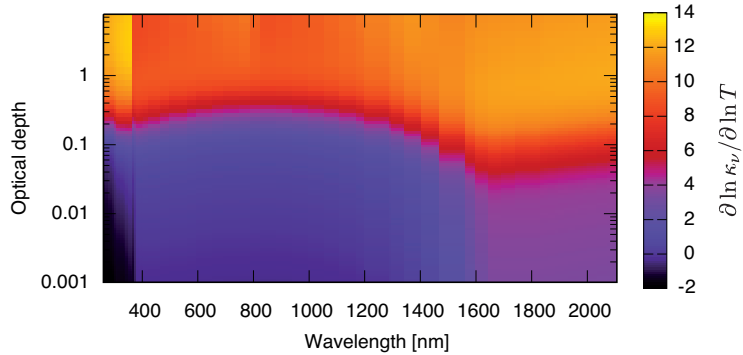}
\caption{Derivative of opacity with respect to temperature as a function of wavelength and the monochromatic optical depth $\tau_\nu$.}
\label{Fig:dlnChi_dlnT} 
\end{figure}

In Figure \ref{Fig:dlnChi_dlnT} the derivative of the opacity with respect to temperature
is shown as a function of wavelength $\lambda$ and monochromatic optical depth $\tau_\nu$.
Although the model atmosphere extends from the photosphere to the transition region, we restrict ourselves to the photosphere since we are only interested in continuum radiation. Certain discontinuities in the plot are noticeable. The corresponding wavelengths coincide with the ionization wavelengths of neutral hydrogen, namely the Balmer continuum ($\lambda=364.6\,$nm). Just above the Balmer continuum there is another ionization edge visible that belongs to magnesium with $\lambda=375.6\,$nm. These discontinuities can be explained with the ionization setting in abruptly at wavelengths just smaller than the ionization edges. If we assume  positive temperature disturbances, then higher temperatures give rise to a higher ionization rate which causes a greater increase of the opacity. Thus, the derivative of the opacity is greater for wavelengths shorter than the ionization edges. The effect diminishes with smaller optical depth, since the influence of neutral hydrogen on the opacity vanishes for higher parts of the atmosphere \cite{vernazza76}. The influence of neutral hydrogen causes the steep rise of the derivative for the deeper layers. For higher layers, the derivative is dominated by the influence of H$^-$-ions \cite{vernazza76} and the opacity is less sensitive to temperature disturbances. This can be explained with the combined effect of the destruction of H$^-$ ions due to higher temperatures and the higher ionization of neutral hydrogen which results in a higher electron density. Investigations showed that if those higher electron densities are not taken into account the derivatives turn out to be negative for those parts of the spectrum where the opacity is dominated by H$^-$. In conclusion, the creation of H$^-$ due to a higher electron density wins over the destruction of H$^-$ due to higher temperatures and thus the opacity increases. There is also a change of the behavior of the derivative for $\lambda = 1645\,$nm. This can be explained by the fact that above this wavelength the free-free processes of H$^-$ starts to dominate the contribution to the opacity.

\begin{figure}
\centering
\includegraphics[width = 0.8\textwidth]{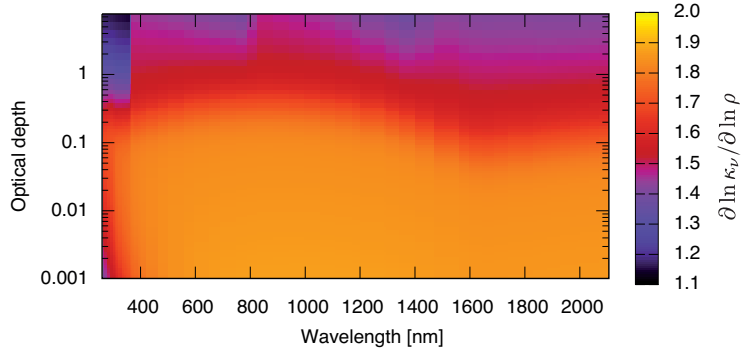}
\caption{Derivative of opacity with respect to density as a function of wavelength and optical depth.}
\label{Fig:dlnChi_dlnRho} 
\end{figure}

The results for the derivative of the opacity with respect to the density are shown in Figure \ref{Fig:dlnChi_dlnRho}.
First, we can point out the discontinuities at the continuum edges again, where the same explanation as for Figure \ref{Fig:dlnChi_dlnT} applies. For deeper layers the discontinuities for the Paschen continuum ($\lambda=820.4\,$nm) and the Brackett continuum ($\lambda=1458.4\,$nm) of neutral hydrogen are more pronounced than for the derivative of the opacity with respect to temperature.

Knowing the derivatives of opacity, it is now possible to calculate the darkening and visibility functions for the most dominant solar oscillation mode with $\ell=0$.

\subsection{Darkening and Visibility Functions}
\begin{figure}
\centering
\includegraphics[width = 0.75\textwidth]{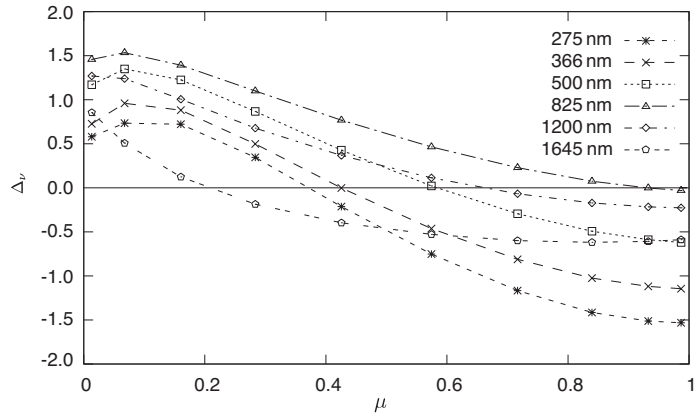}
\caption{The relative darkening functions ($\Delta_\nu$) for selected wavelengths {\it versus} position on the solar disk ($\mu=\cos\theta$). The marks show the sampling points on the solar disk.}
\label{Fig:DarkFunc}
\end{figure}

In Figure \ref{Fig:DarkFunc} the darkening functions ($\Delta_\nu$) (see Section \ref{Sec:Theory}) are shown for selected wavelengths.
Some of the these wavelengths coincide with the wavelengths examined by \inlinecite{zhugzhda96}
and a comparison\footnote{In \inlinecite{zhugzhda96} the graph for $\lambda = 275\,$nm and
$\lambda = 1645\,$nm should be switched.} of the results for those wavelengths shows that
they are similar, although there are some small differences in the position of the
change of sign. This might be caused by a small difference in the derivatives of the opacity derived in this paper and the derivatives used by \inlinecite{zhugzhda96} for higher layers of the atmosphere.  
The wavelengths displayed show that the position of the change of sign on the solar disk
depends strongly on the wavelength. For the wavelengths whose opacity is mostly dominated
by the H$^-$-ions, {\it e.g.} $825\,$nm, the position of the change of sign of the darkening function is almost at
the center of the solar disk while the function increases towards the limb of the Sun.
Thus, for this particular wavelength the oscillation mode $\ell=0$ would be best observed
at the limb while it almost vanishes at the center of the disk. In order to get the
visibility of global oscillation modes for each wavelength, the darkening functions have
to be integrated over the solar disk. This can be done by either numerical integration
of the darkening function for each wavelength or by direct evaluation of Equation
(\ref{Eq:relVisibil}). We choose the latter option for the calculation of the monochromatic
visibility function.

\begin{figure}
\centering
\includegraphics[width = 0.75\textwidth]{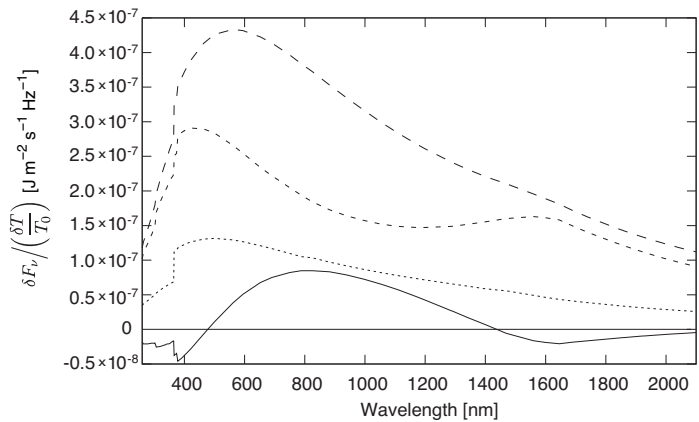}
\caption{The absolute visibility function versus wavelength (solid line). In
addition, the different contributions $\partial B_\nu/\partial \ln T$ (long dash), $\partial \ln \kappa_\nu / \partial \ln T$ (medium dash), and $\partial \ln \kappa_\nu / \partial \ln \rho$ (short dash) originating from the Planck function and the opacity are shown.}
\label{Fig:absVisibil}
\end{figure}

The result of this integration can be seen in Figure \ref{Fig:absVisibil}, where the
absolute visibility function ($\delta F_\nu/(\delta T/T_0)$) is plotted together with
its different contributions from the Planck function and opacity.
This plot clearly indicates, that the visibility functions are
dominated by the effects of the opacity. This is due to the
assumption of adiabatic oscillations for which the temperature and
density disturbances are in phase. The effect of the oscillation
on the opacity is big enough in order to overcome the disturbances
of the Planck function and for shorter and longer wavelengths
positive disturbances actually result in a darkening of the solar
flux. Considering the visibility function, the oscillation mode
$\ell=0$ has the biggest influence on the disturbance of the flux at wavelengths around $\lambda
= 800\,$nm. Since the visibility functions also show a
change of sign for two wavelengths of the considered spectrum, it
is interesting to note that for $\lambda = 480\,$nm and $\lambda =
1450\,$nm the global oscillation of mode $\ell=0$ should not be
visible. This can actually be used as a test for the assumption of
adiabatic oscillations as will be discussed below.

\section{Conclusions}

There is  an essential distinction between visibility functions
for the global solar and stellar oscillations observed by
measurements of Doppler shifts and continuum brightness
fluctuations. In the case of Doppler shifts the visibility is
defined by the projection of the velocity vector of the oscillation
modes onto the line of sight. Thus, it is not complicated to calculate
the visibility function for every spherical harmonic.

 The derivation of the visibility functions for brightness
fluctuations due to global oscillations is more difficult. We have succeeded in
obtaining the spectral darkening and visibility functions over a
broad range of the visible and infrared spectrum of the Sun. Two
simplifications of the problem make this possible: the
adiabatic approximation and the condition $\delta T/T_0=\mathrm{constant}$.

We found that the brightness fluctuations are not necessarily
proportional to temperature fluctuations. The darkening function
shows a change of sign and the visibility function can be
negative, {\it e.g.} the brightness can decrease with the increase
of the temperature due to radiative flux blocking.

But there are some problems considering the visibility function.
In accordance with Figure \ref{Fig:absVisibil}, the brightness
fluctuations should be absent for $480\,$nm and $1450\,$nm. But
this is not the case. Brightness fluctuations are observed at
those wavelengths by SOHO/VIRGO and CORONAS/DIFOS. This clearly
indicates that one of the assumptions is not correct, if not
both of them. The condition with constant temperature disturbances
should be relaxed in general. It is necessary to know the temperature
fluctuations in dependence of the depth in the solar atmosphere.

The second assumption in use is the adiabatic approximation. Our
calculations confirm one of the results of \inlinecite{toutain93}.
If adiabatic conditions are assumed, the opacity perturbations
show a great influence on the visibility functions. But as
mentioned above, in this case the brightness fluctuations should
not be visible for $480\,$nm and $1450\,$nm which is not in
accordance with observations. Thus, this approximation is violated
in the solar photosphere. \inlinecite{zhugzhda06} discovered that
there are phase shifts between brightness fluctuations in
different spectral channels of the photometric instrument DIFOS.
This is only possible if oscillations are nonadiabatic since in
this case there is a phase shift between the temperature and
density fluctuations. Besides, the  nonadiabatic {\it p} modes are
coupled with temperature waves \cite{zhugzhda83}. Consequently,
the darkening and visibility functions for nonadiabatic
oscillations are complex functions. Multichannel observations of
CORONAS together with SOHO observations make it possible to
explore these nonadiabatic oscillations.

The visibility and darkening functions are needed for the
interpretation of the observations of solar and stellar
oscillations. The main result of the current exploration is that
the effect of the opacity fluctuations on brightness oscillations
dominates. But it is a rather complicated problem to find out the
correct functions for adiabatic and nonadiabatic oscillations. The
current paper just presents the first step on this road.

\acknowledgements YZ is thankful to RFFI (grant 06-02-16359). CN, MR, YZ and OvdL acknowledge support from the European Helio- and Asteroseismology Network -- HELAS. HELAS is funded by the European Union's Sixth Framework programme. We also thank J. Staude for his helpful comments on our research. We thank H. Uitenbroek for the provision of his radiative transport code RH. Furthermore, we gratefully thank the anonymous referee for their substantial revision and valuable advice that helped to improve this paper.

\bibliographystyle{spr-mp-sola.bst}
\bibliography{Bib_Helioseismology}
\end{article}

\end{document}